\documentclass{aa}  

\usepackage[breaklinks=true]{hyperref}
\usepackage{graphicx}

\usepackage{caption}

\usepackage{amssymb}

\usepackage{enumerate}

\usepackage{subfigure}

\usepackage[toc,page]{appendix}

\usepackage[multiple]{footmisc}

\usepackage{booktabs}

\bibpunct{(}{)}{;}{a}{}{,}

\usepackage{natbib,twoopt}

\def\aj{AJ}%
\def\araa{ARA\&A}%
\def\apj{ApJ}%
\def\apjl{ApJ}%
\def\apjs{ApJS}%
\def\ao{Appl.~Opt.}%
\def\aap{A\&A}%
\def\aaps{A\&AS}%
\def\mnras{MNRAS}%
\def\pasp{PALMP}%
\def\solphys{Sol.~Phys.}%
\def\nat{Nature}%
\def\gca{Geochim.~Cosmochim.~Acta}%
\def\physscr{Phys.~Scr}%
\begin{document} 

   \title{CNO behaviour in planet-harbouring stars}
   
   \subtitle{II. Carbon abundances in stars with and without planets using the CH band\thanks{Based on observations collected at the La Silla Paranal Observatory, ESO (Chile) with the HARPS spectrograph at the 3.6-m telescope (ESO runs ID 72.C-0488, 082.C-0212, and 085.C-0063)}\thanks{Table 1 is only available in electronic form at the CDS via anonymous ftp to cdsarc.u-strasbg.fr (130.79.128.5)
   		or via http://cdsweb.u-strasbg.fr/cgi-bin/qcat?J/A+A/}}

   \author{L. Su\'arez-Andr\'es 
          \inst{1,2}
                        , G. Israelian    \inst{1,2}     
                        , J.I. Gonz\'alez Hern\'andez \inst{1,2}
                        , V. Zh. Adibekyan \inst{3}
                        , E. Delgado Mena \inst{3}
                        , N. C. Santos \inst{3,4}
                \and
                        , S. G. Sousa \inst{3,4}
          }

   \institute{Instituto de Astrof\'isica de Canarias, E-38205 La Laguna,
Tenerife, Spain \newline \email{lsuarez@iac.es}
              \and
           Depto. Astrof\'isica, Universidad de La Laguna (ULL),
E-38206 La Laguna, Tenerife, Spain
        \and
          Instituto de Astrof\'isica e Ci\^encias do Espa\c{c}o, Universidade do Porto, CAUP, Rua das Estrelas, 4150-762 Porto, Portugal 
         \and Departamento de F\'isica e Astronomia, Faculdade de C\^iencias, Universidade do Porto, 4169-007 Porto, Portugal} 

   \date{Received August, 2016; accepted November, 2016}

 
  \abstract
  {Carbon, oxygen and nitrogen (CNO) are key elements in stellar formation and evolution, 
and their abundances should also have a significant impact on planetary formation and evolution.}
   {We aim to present a detailed spectroscopic analysis of 1110 solar-type stars, 143 of which are known 
to have planetary companions. We have determined the carbon abundances of these stars and investigate a 
possible connection between C and the presence of planetary companions.}
  {We used the HARPS spectrograph to obtain high-resolution optical spectra  of our targets. 
Spectral synthesis of the CH band at 4300\AA \space was performed with the spectral synthesis codes MOOG and FITTING.}
{We have studied carbon in several reliable spectral windows and have obtained abundances and 
distributions that show that planet host stars are carbon rich when compared to single 
stars, a signature caused by the known metal-rich nature of stars with planets. We find no 
different behaviour when separating the stars by the mass of the planetary companion}
   {We conclude that reliable carbon abundances can be derived for solar-type stars from the CH 
band at 4300\AA. We confirm two different slope trends for [C/Fe] with [Fe/H] because the behaviour is
 opposite for stars above and below solar values. We observe a flat distribution of the [C/Fe] ratio for all 
planetary masses, a finding that apparently excludes any clear connection between the [C/Fe] abundance ratio and planetary mass.}
 

\titlerunning{Carbon abundances in planet-harbouring stars.}
\authorrunning{Su\'arez-Andr\'es, L. et al.}
%
\maketitle
%
\section{Introduction}
Carbon, together with nitrogen and oxygen,  is one of the most abundant elements, after H and He. 
These elements play an important role in the lifetime of solar-type stars as they generate energy through the CNO cycle \citep{liang}

Carbon is created through a different process from nitrogen and oxygen. In this case, $\alpha$-chain 
reactions are the dominant processes. Both massive ({M${_\star}$} > 8$M_{\odot}$) and low- to 
intermediate-mass stars contribute to carbon production, but the relative 
importance of each type of star has not been clarified owing to uncertainties in the metallicity-dependent mass loss \citep[see ][]{meynet02, hoeck97}

It has been well-established that stars hosting a giant planet are, on average, more metal-rich 
than stars with no detected companion  \citep{santos01,santos04,fischer05,gonzalez06, buchhave}. 
Two hypotheses have been proposed in an attempt to explain the  nature of these metal-rich stars: The first one is self-enrichment, where the host star is polluted by planetesimals accreting on to the star. \citep[see][]{israelian01, israelian03}
This self-enrichment scenario could lead to a relative 
overabundance of refractories, such as Si, Mg, Ca, Ti, and the iron-group elements, compared to 
volatiles, such as C, N, O, S, and Zn \citep{gonzalez97, smith01}. 
The second one is the existence of a primordial cloud. \citet{santos01, santos02, santos03} found evidence
that this over-abundance most is probably caused by a metal-rich primordial cloud.

The analysis of the HARPS stellar sample by \cite{adi12} shows that this chemical overabundance found in stars with planets is not exclusive to iron.
 Using a sample of 1111 FGK stars, they show that stars hosting a giant planet in their systems show an over-abundance 
in 12 refractory elements. They find that, at metallicities below $-$0.3 dex, about 76\% of the planet 
hosts are enhanced in $\alpha$ elements at a given metallicity. This proves that planet formation requires a 
certain amount of metals, even for low-mass planets \citep{adi12c}.

There are several studies of carbon in solar-type stars, but hardly any of these use CH molecular features \citep[see][]{clegg}.
 Instead, the most commonly used features are the $C_{2}$ Swan ($\lambda$ 5128 and $\lambda$5165), the CN molecular
 band ($\lambda$4215), and two atomic lines ($\lambda$ 5380.3 and $\lambda$5052.2). This last feature becomes 
very weak for cool stars ($\rm T_{\rm eff}$ < 5100K). 

Previous studies by \citet{elisa10} and \citet{dasilva11, dasilva15}
 using some of the above-mentioned carbon features show that there is no difference in behaviour regarding carbon between stars with and without any 
known planetary companion. The enrichment found was attributed to Galactic chemical evolution.

In this article we study the volatile C element using the CH band and derive C abundances for 1110 HARPS FGK 
stars, thereby complementing the \cite{adi12} study of refractory elements. We propose that CH molecular features are a good alternative to atomic lines.
We also investigate whether the presence of a (giant) planetary companion affects the carbon abundance of the planet host stars.

\section{Sample description}

The high-resolution spectra analysed in this work were obtained with the HARPS spectrograph  at La Silla Observatory
(ESO, Chile) during the HARPS GTO programme
 \citep[see][for more information about the programme]{mayor03, locurto10, santos11}.
The spectra had already been used in the analysis of stellar parameters, as
 well as the derivation of precise chemical abundances \citep[see e.g.][]{sousa08,sousa11a, sousa11b, tsantaki, adi12}.

The high spectral resolving power ($R=110,000$) of HARPS and a good signal-to-noise ratio (S/N) are an
 optimal combination for properly analysing the CH band at 4300\AA. About 41$\%$ of the stars within our sample, 
 at 4300\AA, {have} S/N>150, 18$\%$ have S/N in the range 100--150, and 30$\%$ have S/N below 100. Only 4$\%$ of our sample have S/N lower than 40.

The sample consists of 1110 FGK solar-type stars with effective temperatures between 4400 K
 and 7212 K, metallicities from $-1.39$ to $0.55$ dex, and surface gravities from $3.59$ to 
$4.96$ dex. 143 out of 1110 are planet hosts,\footnote{Data from www.exoplanet.eu.} whereas the 
other 967 are {comparison sample} or comparison stars (stars with no known planetary companion).

\section{Analysis}
\subsection{Stellar parameters and chemical abundances}
The stellar parameters used in this study were taken from \citet{sousa08,sousa11a, sousa11b} 
and \citet{tsantaki}, who used the same spectra as we did for this study. All the stellar parameters were 
derived by measuring equivalent widths of Fe{\rm I} and Fe{\rm II} lines using the ARES and ARES2 code 
\citep{ares,ares2}. Chemical abundances of elements with spectral lines present in the  features studied
were obtained from \cite{adi12}. These elements are Ca, Si, Sc, and Ti.
Also, chemical abundances of elements other than carbon were adapted in targets with more recent stellar
parameters (also obtained by our group using the aforementioned technique, see \citealt{tsantaki}) following 
uncertainties presented in their original sources, so systematic effects in the chemical abundances or the
stellar parameters are negligible and {do not} affect the consistency of our results.

Carbon abundances were determined using a standard local thermodynamic equilibrium (LTE) 
analysis with the MOOG spectral synthesis code  \citep[2013 version]{sneden} and a grid of Kurucz 
(1993) ATLAS9 atmospheres. All atmospheric parameters, $T_{\rm eff}$, $\log g$, $\xi_{t}$, and 
[Fe/H] were taken, as mentioned above, from \citet{sousa08, sousa11a, sousa11b} and \citet{tsantaki}. 
Adopted solar abundances for carbon and iron were $\log\epsilon(\rm C)_{\odot}=8.50$ dex 
and $\log\epsilon(\rm Fe)_{\odot}=7.52$ \citep{caffau, caffau11}. For oxygen, we adopt $\log\epsilon(\rm O)_{\odot}=8.71$ dex \citep{sara}.

  As molecular carbon abundances are affected by molecular equilibrium, we need to consider oxygen and nitrogen. 
  Oxygen, being an $\alpha$-element, does not keep pace 
  with [Fe/H] as nitrogen does \citep[see][]{sara,nitrogeno}.  However, there were 575 stars with no previous
        measurements of O and 1054 stars with no previous measurements of N. We decided to use the available oxygen 
        $\lambda6158$ results and perform an interpolation for these 575 stars with no oxygen measurements  (see 
        Section 4.1). We obtained individual O and N 
        abundances for the available targets from \cite{sara} and \cite{nitrogeno}, respectively.
   As nitrogen re-scales with [Fe/H], we obtained carbon abundances, thus providing individual values of oxygen and scaling 
   nitrogen with the metallicity.

 We did not consider NLTE effects because these features are likely to be immune to 
 such effects at these metallicities, although little computational work has been done in this area \citep{asplund05, schuler08}.
  
\subsection{CH band}

The CH band is the strongest feature observed in the spectral region $\lambda 4300\AA.$ We 
determined carbon abundances by fitting synthetic spectra to data in this wavelength range. The dissociation potential
 used for CH spectra is $D_{o}=3.464$ eV, as recommended in \citet{grevesse90}. The complete line list used in this work 
was obtained from VALD3 \citep{vald}. We slightly modified  the log $gf$ values of the strongest lines in order to fit the solar spectrum.

         The continuum was normalised locally with a fifth-degree polynomial, using the CONTINUUM task of IRAF.{\footnote{IRAF is 
         distributed by the National Optical Astronomy Observatory, which is operated by the Association of Universities for
         Research in Astronomy (AURA) under a cooperative agreement with the National Science Foundation.}}

To find the best fit abundance value for each star, we used 
the FITTING program \citep{jonay11} and the MOOG synthesis  code in 
its 2013 version. The best fit was obtained using a $\chi^{2}$ minimization procedure, by comparing each 
synthetic spectrum with the observed one in the following spectral regions:  $\lambda\lambda4276.8-4282.2\AA$,
$\lambda\lambda4292.8-4293.4\AA$,
$\lambda\lambda4294.5-4298.3\AA$,
$\lambda\lambda4301.5-4303.7\AA$, 
$\lambda\lambda4307.1-4310.7\AA$, 
$\lambda\lambda4322.6-4324.6\AA$, 
$\lambda\lambda4360.1-4360.6\AA$, 
$\lambda\lambda4362.3-4367.0\AA$, 
$\lambda\lambda4377.0-4377.4\AA$,
and $\lambda\lambda4387.2-4387.7\AA$.
These regions were chosen for the presence of relatively strong CH features that allow reliable abundance measurements.
We used a $\chi^{2}$ comparison between observed and synthetic spectra, and we define $\chi^{2}$=$\sum_{i=1}^{N}(F_{i}-S_{i})^{2} / N $,
 where $F_{i}$ and $S_{i}$ are the observed and synthetic fluxes, respectively, at wavelength point $i$. Best-fit carbon abundances
 were extracted from every spectral range and then the final carbon abundance for each star was computed as the average of these values.
 {All the atmospheric parameters, 
 $\rm T_{\rm eff}$, $log g$, $\xi_{t}$, [Fe/H] as well as all the chemical abundances used
 in the determination of C (see Section 3.1 and 4.1), were
 fixed. Rotational broadening was set as a free parameter with $v$ sin$i$ varying between 0.0 and 14.0 $km/s$ with a step of 
 1 km/s. For more information about the procedure followed, see \cite{nitrogeno}.}

\begin{figure}[!h]

        \centering
                        \scalebox{1.3}[1.1]{
                    \includegraphics[angle=180,width=0.75\linewidth]{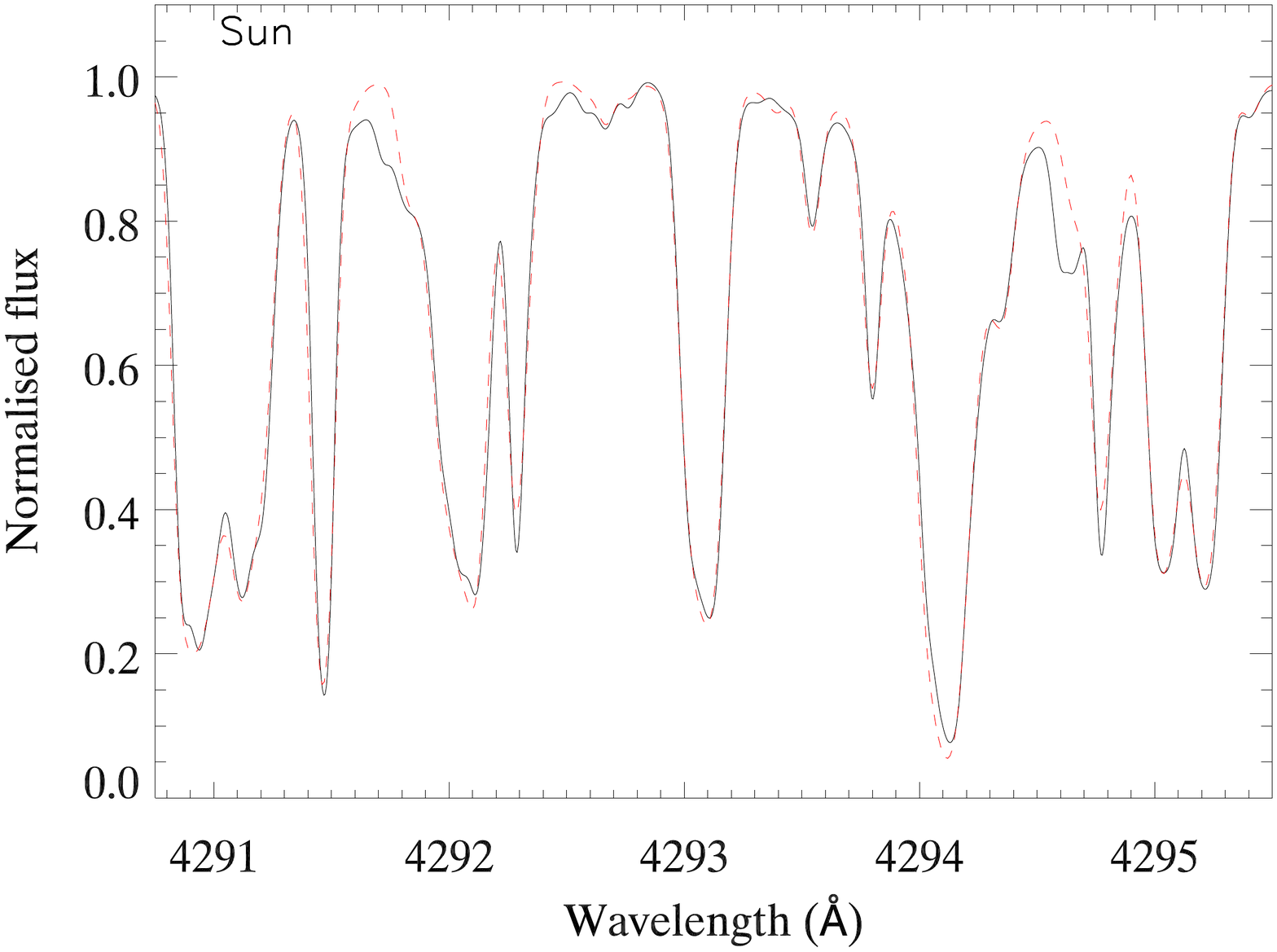}}
                        \scalebox{1.3}[1.1]{
                        \includegraphics[angle=180,width=0.75\linewidth]{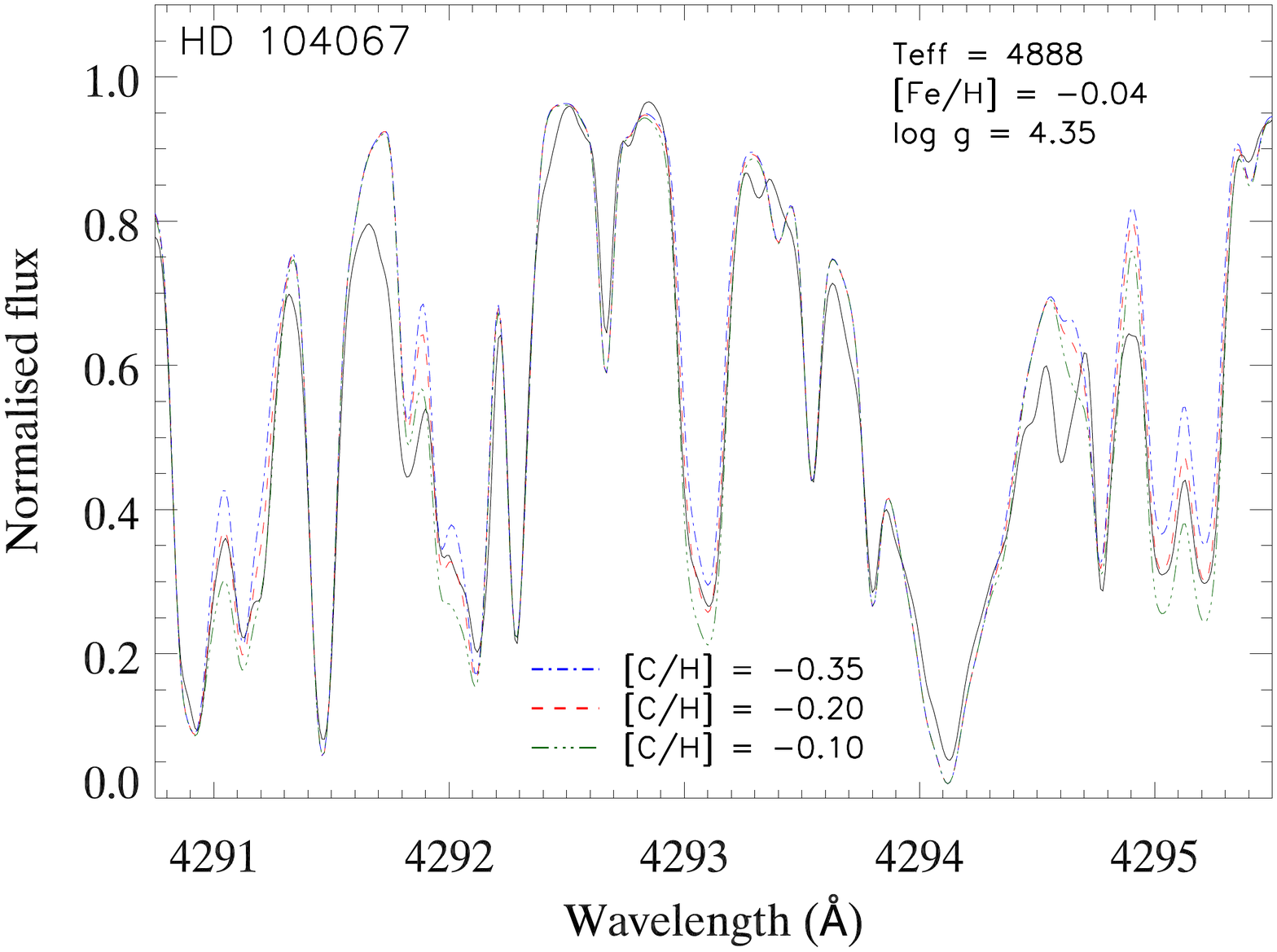}}
                                \scalebox{1.3}[1.1]{
                                        \includegraphics[angle=180,width=0.75\linewidth]{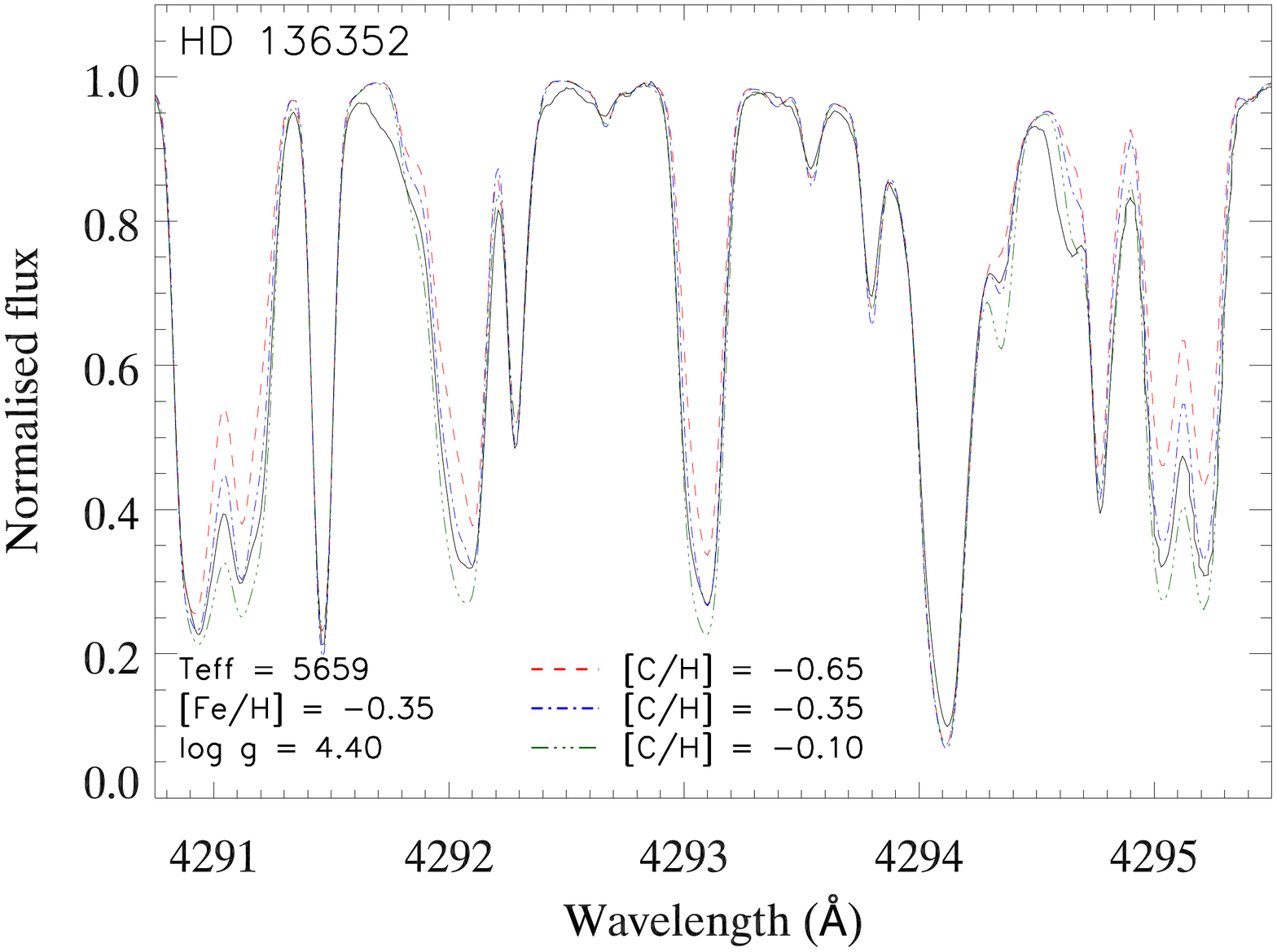}}

        \caption{Top panel: observed  solar spectrum (solid) and synthetic spectra (red-dashed lined). 
{Middle and bottom panel}: observed (solid) and synthetic spectra (green-dotted lined, red-dashed, 
and blue-dashed dotted) of HD 104067 and HD 136352.}
        \label{Fig:fit1}

\end{figure}

In Fig.~\ref{Fig:fit1} we show the observed and synthetic spectra for the Sun and two stars that are {depicted} for different 
temperature and metallicity within our sample. For these two stars, the best fit and two  different carbon abundances are also shown.

\begin{figure}
        
        \centering
                        \scalebox{1.3}[1.2]{
                                \includegraphics[angle=180,width=0.8\linewidth]{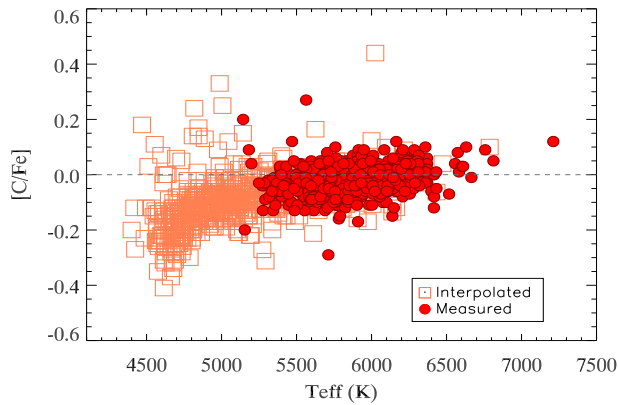}}
        \caption{[C/Fe] as a function of $\rm T_{\rm eff}$. Filled red {circles} indicate those stars 
with available {oxygen} information, and open {red squares} indicate those stars with interpolated oxygen values.}
        \label{inter}
\end{figure}

To examine how variations in the atmospheric parameters affect carbon abundances, we tested the [C/H] sensitivity 
in stars with widely differing parameters, given the wide range of stellar parameters in our sample. For each set of 
stars, we tested the carbon abundance to changes in the stellar parameters ($\pm$50 K for $T_{\rm eff}$ and 
$\pm$0.1 dex in $\log g$ and metallicity). Results are shown in  Table \ref{sens}. The effect of microturbulence 
was not taken into account because an increase of 0.1 km s$^{-1}$ produced an average decrease of 0.0005 dex in carbon 
abundances, which is negligible with regard to the effects of other parameters. An error due to a continuum placement was 
considered for all stars of 0.05 in those stars with higher S/N and 0.1 for those with lower S/N, thereby increasing the 
considered continuum error with decreasing S/N. All effects were added quadratically to obtain the final uncertainties in 
carbon abundances following the expression:

\scalebox{0.85}{
$\Delta {\rm [C/H]} = (\Delta_\sigma^2+\Delta_{T_{\rm eff}}^2+\Delta_{\log g}^2+\Delta_{\rm [Fe/H]}^2+\Delta_{\rm cont}^2+\Delta_{\rm interp}^2)^{1/2}$,}

\bigskip
{where $\Delta_{\sigma}$ refers to the error due to the $\chi^{2}$-fitting.} The term $\Delta_{\rm interp}^2$ is added only for those stars with interpolated oxygen values.

\section{Results}

\subsection{Carbon abundances}

In 2015, Bertr\'an de Lis et al.\ (using the same spectra 
as in this work) obtained oxygen abundances for 698 stars by studying two atomic features: 
$\lambda6158$ and $\lambda6300$. We do not 
use $\lambda6300$ results,    because of the uncertainties of that feature caused by the presence of a
blended nickel line. In order to obtain oxygen values for all our sample, we performed an interpolation of 
their [O/Fe] vs [Fe/H] results only for the $\lambda6158$ feature. {The interpolation consist of a 3$^{rd}$ degree polynomial 
($ax^{3}+bx^{2}+cx+d)$, where the coefficients are: $a=0.33$,$b=0.75$, $c=-0.36$, $d=0.03$).
All the interpolations are consistent with an [O/H] vs [Fe/H] plot with previous measurements  \citep{jonay13}.}
 We obtained oxygen abundances for 575 stars in order to complete the  sample of 1110 {stars.}

{ In Fig. \ref{inter} we  see carbon abundances as a function of $T_{\rm eff}$. We can see that for}
 $T_{\rm eff} > 4800$ K, both sets of data, those using real data and those using interpolated data, 
 follow the same increasing trend for $T_{\rm eff}$ with similar dispersion 
(all stars from 4400 K to 5144 K have interpolated oxygen data). For those stars with $T_{\rm eff} < 4800$ K
 we find a steep decrease in [C/Fe] abundance ratios with temperature. A similar effect can also be found for other
 $\alpha$-elements \citep{adi12}, as well as for nitrogen \citep{nitrogeno}.
 
 Stars with effective temperatures below 4800 K were excluded from the analysis because of uncertainties in the behaviour of those cool 
 stars: we find no explanation for the decrease in carbon abundance as we move to lower temperatures.

\subsection{Comparison with the literature}

\begin{figure}[!hbtp]
        \begin{center}
                \scalebox{1.3}[1.2]{
                        \includegraphics[angle=180,width=0.8\linewidth]{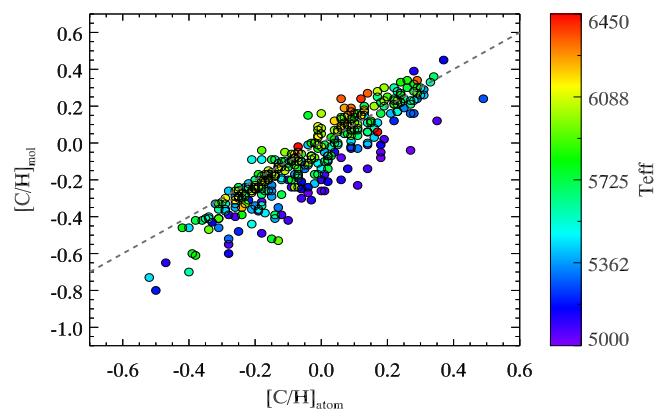}}
                                \scalebox{1.3}[1.2]{
                                        \includegraphics[angle=180,width=0.8\linewidth]{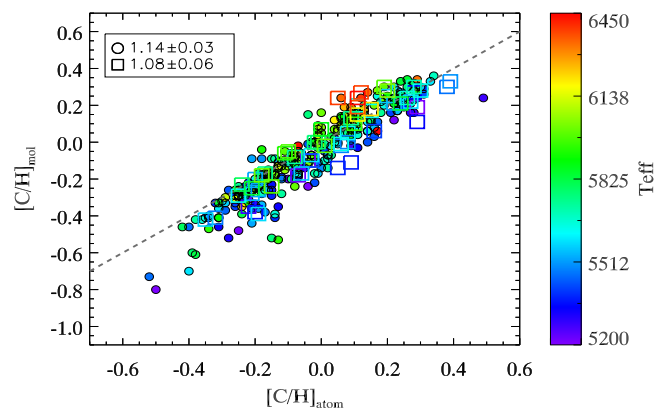}}
                \caption{Top panel: molecular [C/H] results {from this work} as a 
                function of atomic [C/H] extracted from \cite{elisa10}. The dashed line represents 
                a 1:1 relation. Bottom panel: comparison for common stars with the works of
                \cite{elisa10} {(filled circles)} and \cite{nissen14} {(open squares)} for stars with $\rm T_{\rm eff}$>5200K.}
                
                \label{comp}
        \end{center}
\end{figure}

A few studies can be found that use the CH band at 4300\AA, as opposed to C atomic lines. 
\cite{elisa10} studied, among other elements, the carbon abundances of 451 stars of the HARPS sample, 
using the same spectra that we did. They used two atomic lines: $\lambda5380.3$ and $\lambda5052.2$. 
To confirm the validity of our study, we will compare our results of stars in common. In Fig. \ref{comp}{, top panel,}
we can see the molecular [C/H] results of these stars as a function of atomic [C/H]. 
Representing the 1:1 relation with a dashed line, we can see how the results show good agreement 
between themselves for effective temperatures above 5200 K, whereas for cooler stars the results show some disagreements. 
{Thus, we have decided to continue only with the analysis of stars with effective temperatures above 5200 K.}

{In the bottom panel of Fig. \ref{comp} we show two sets: stars with $\rm T_{\rm eff}$>5200 in common  
        with \cite{elisa10} (filled circle)  and
        stars with $\rm T_{\rm eff}$>5200 in common with \cite{nissen14} (open squares). Carbon abundances
        from \cite{nissen14} were obtained by studying {the same atomic lines as in \cite{elisa10}}. The slope of the fit, with
        its standard deviation, is provided for each set. As
        can be seen in Fig. \ref{comp}, bottom panel, all sets (data
        from this work, \cite{elisa10} and \cite{nissen14}) show  good agreement}

\section{Carbon abundances in stars with and without planets}

We analysed optical high-resolution spectra of 112 planet host stars and 639 comparison stars. We aim to
explore possible differences in carbon abundances between both 
samples. Our results can be found in an online table, along with the stellar parameters. A sample listing
 is shown in Table \ref{tablacon}. \footnote{Full Table 1 is  available online.}

\begin{table}[!hbtp]\scriptsize
        \caption{Sample of carbon abundances for a set of stars with and without planets (see online table).}
        \label{tablacon}
        \centering
        \scalebox{0.75}[1]{
        \begin{tabular}[l c c c c c c c]{ l  c  c c c c c c}
                \hline
                \hline
                Star    &       $\rm T_{\rm eff}$       &       $log g$  & $\xi_{t} $    &       [Fe/H]  &       [C/H]  & Planet? & Pop.\\  
                
                &       (K)     &       (cm $s^{-2})$ &  $(km s^{-1})$  &               &       & &\\  \hline

                HD 114386 & 4774 &  4.37 & 0.01 &-0.09 & -0.11 $\pm$ 0.11 & Y & Thin \\
                HD 104067 & 4888 &  4.35 & 0.66 & -0.04 & -0.16 $\pm$ 0.09 & Y & Thin \\
                HD 169830 & 6370 & 4.20 & 1.56 & 0.18 & 0.24 $\pm$ 0.08 &Y & Thin\\
                HD 136352 & 5659 & 4.40 & 0.90 &  -0.35 & -0.35 $\pm$ 0.07 &Y& Thick\\
                
                BD+063077 & 6136 & 4.95 & 1.07 & -0.36 &-0.33 $\pm$ 0.15 &N & Thick\\
                HD 215902 & 5454 &  4.46 & 0.53 & -0.25 & -0.32 $\pm$ 0.12 &N & Thin\\
                HD 125522 & 4839 &  4.45 & 0.40 &-0.46 & -0.45 $\pm$ 0.12 &N & Thin\\
                HD 129229 & 5872 &  3.89 & 1.37 & -0.42 & -0.42 $\pm$ 0.09 &N & Thin\\
                
                ... &    &   &    &  &   & & \\
                \hline \\
        \end{tabular}
}
        
        
\end{table}

{To test for a possible relation between carbon abundances and the masses of planetary companions, 
we separated the planet population into two groups: low-mass planets (LMP; with masses less than or equal
to 30 $M_{\oplus}$) and high-mass planets (HMP; with masses greater than 30 $M_{\oplus}$). In those
stars which host several planets, the most massive planet was considered in our study. {Our sample consists of
23 low-mass and 89 high-mass planet hosts.}}

\begin{figure}[!hbtp]
        \begin{center}
                \scalebox{1.3}[1.2]{
                        \includegraphics[angle=180,width=0.8\linewidth]{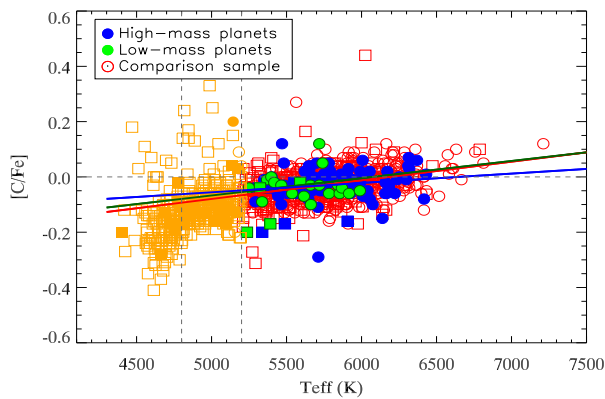}}
        \end{center}
        \caption{[C/Fe] vs. $T_{\rm eff}$. Filled circles represent planet hosts and open circles, the comparison sample. Squares represent interpolated oxygen values.}
        \label{teff}
\end{figure}

We also look for distinguishable trends between these samples by representing [C/Fe] as a function of
 $T_{\rm eff}$ (see Fig. \ref{teff}). As mentioned, we find a increasing
 trend of the [C/Fe] abundances on $T_{\rm eff}$ (for both planet and non-planet-hosts).
 {However, the spread of [C/Fe] abundance values with $T_{\rm eff}$ amounts to about 0.1 dex from
 5200~K to 6200~K, with $\sim$65\% of our sample with [C/Fe] < 0.0. Statistics for these 
 relations are provided in Table \ref{table:stats1}. } In Fig. \ref{teff} we can see the line at 4800 and 5200 K separating the cool stars from the  sample studied.

\begin{figure}
        
        \centering
        
        \scalebox{1.3}[1.2]{
                \includegraphics[angle=180,width=0.8\linewidth]{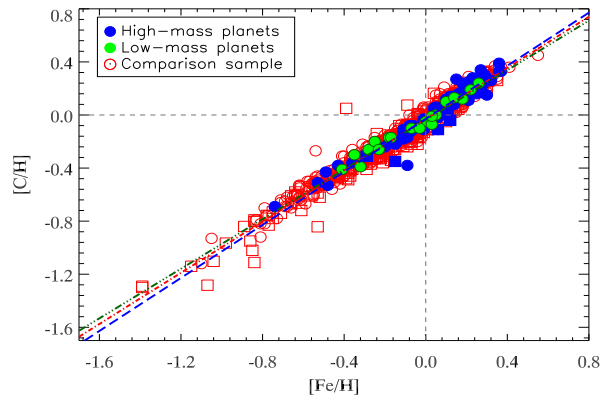}}
        
        \scalebox{1.3}[1.2]{
                \includegraphics[angle=180,width=0.8\linewidth]{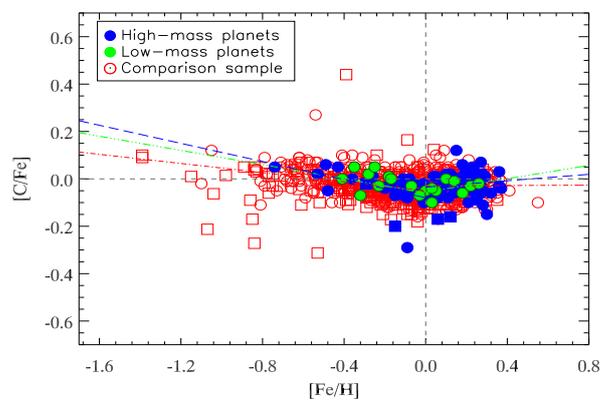}}
        
        \caption{[C/H] and [C/Fe] vs.\ [Fe/H] plots. Filled circles represent planet hosts and the open 
circle, the comparison sample. Dotted lines represent solar values. Squares stand for interpolated oxygen values.}
        \label{Fig:met1}
        
\end{figure}

We looked for distinguishable trends between the host and {non-planet host} samples by representing
 [C/Fe] and [C/H] abundance ratios as functions of [Fe/H] (see Fig.~\ref{Fig:met1}). In 
Fig. \ref{Fig:met1} top panel, [C/H] vs. [Fe/H] is represented. We can see how {all} the samples 
behave in the same way, and that {no difference between them can be found}. If we look at the slopes
 of each sample, we find very similar values, thus confirming that stars with and without planets 
follow the same trend.

In the bottom panel of Fig. \ref{Fig:met1}, we find that [C/Fe] follows the same trend found in other $\alpha$-elements. 
As presented in \citet{dasilva15}, a decreasing trend can be found for metallicity values below solar,
whereas for super-solar values, this trend has a positive slope. We can see in Fig.~\ref{Fig:met1} 
how the slope for planets and the {comparison sample} at high metallicities follows the same behaviour, whereas for 
the low iron abundances the planet-host trend is steeper than for {the 
comparison sample} (for {comparison sample} only those stars with [Fe/H] > -0.74 are taken 
(the minimum metallicity found for a planet host star in our sample)). {This behaviour is consistent for stars with both low- and high-mass planetary companions}. Statistics for these relations are also provided in Table \ref{table:stats1}.

{To evaluate the significance of the correlations we performed a simple bootstrapped 
Monte Carlo test. For more details on the test we refer the reader to \cite{figueira13} and \cite{adi14}.}

\begin{table}[!t]
        \caption{{Statistics for the three groups studied (Comparison sample - NP, 
        low-mass planets - {LMP}, high-mass planets - HMP) for [C/H] vs [Fe/H] and [C/Fe] vs 
        Teff and [Fe/H] (see Figs. \ref{teff} and \ref{Fig:met1}).}}             
        \label{table:stats1}      
        \centering   
        \scalebox{0.9}[0.97]{      
                \begin{tabular}{  c c c c c } 
                        \hline
                        \hline
                        & & \multicolumn{1}{c}{[C/Fe] vs Teff}\\
                        \cmidrule{2-5} 
                        &     & & & \\
                        &   Slope & Slope$\rm_{\rm err}$ & R\tablefootmark{*} & $\sigma_{R}$ \\
                        \cmidrule{2-5}        
                        NP  & 6.74E-05 & 7.13E-06 & 0.35 & 8.7 \\
                        LMP  & 6.24E-05 & 3.95E-05 & 0.33 & 1.5 \\
                        HMP  & 3.38E-05 & 2.16E-05 & 0.17 &  1.6 \\    
                        \cmidrule{1-5}                    
                        & & \multicolumn{1}{c}{[C/H] vs [Fe/H]}\\
                        \cmidrule{2-5} 
                        &     & & & \\
                        &   Slope & Slope$\rm_{\rm err}$ & R & $\sigma_{R}$ \\
                        \cmidrule{2-5}        
                        NP  & 0.97 & 8.02E-03 & 0.98 & 24.5 \\
                        LMP  & 0.93 & 4.06E-2 & 0.98 & 4.6 \\
                        HMP  & 1.00 & 2.90E-02 & 0.97 &  9.0 \\    
                        \cmidrule{1-5}                    
                        & & \multicolumn{1}{c}{[C/Fe] vs [Fe/H]}\\
                        & & \multicolumn{1}{c}{{\scriptsize {(for [Fe/H] < 0.0)}}}\\
                                                \cmidrule{2-5} 
                                                &     & & & \\
                                                &   Slope & Slope$\rm_{\rm err}$ & R & $\sigma_{R}$ \\
                                        
                                                \cmidrule{2-5}        
                                        
                                                NP  & -0.10 & 1.57E-02 & -0.31 & 6.1 \\
                                                LMP  & -0.15 & 8.24E-02 & -0.46 & 1.7 \\
                                                HMP  & -0.20& 8.31E-02 & -0.49 &  2.5 \\    
                        \cmidrule{1-5}                    
                        & & \multicolumn{1}{c}{[C/Fe] vs [Fe/H]}\\
                        & & \multicolumn{1}{c}{{\scriptsize {(for [Fe/H] > 0.0)}}}\\
                        \cmidrule{2-5} 
                        &     & & & \\
                        &   Slope & Slope$\rm_{\rm err}$ & R & $\sigma_{R}$ \\
                        \cmidrule{2-5}        
                        NP  & 0.00 & 3.46E-02 & 0.01 & 0.1 \\
                        LMP  & 0.14 & 1.10E-1 & 0.44 & 1.2 \\
                        HMP  & 0.07 & 7.50E-02 & 0.12 &  0.9 \\    
                        \cmidrule{1-5}                    

                        \hline                
                \end{tabular}}
                \tablefoot{
                        \tablefoottext{*}{R and $\sigma_{R}$ are the correlation coefficients and their standard deviation, respectively.}}              
                        
        \end{table}

\begin{table*}[!hbtp]
        \caption{Sensitivity of the carbon abundance derived from the CH band at {4300}\AA. Changes of 
50K in $T_{\rm eff}$, 0.1 dex in gravity and 0.1 in [Fe/H] were applied.}             
        \label{sens}      
        \centering   
        \scalebox{1}[1]{      
                \begin{tabular}{ c c c c c} 
                        \hline
                        \hline
                        & & &\multicolumn{1}{c}{Star}\\
                        & & &\multicolumn{1}{c}{($\rm T_{\rm eff}$; $log g$; [Fe/H])}\\
                        \cmidrule{2-5} 
                        &     & & & \\
                        & HD 103720 &  HD 222595 & HD 75881 & HD 66740 \\
                        & (5017; 4.43; -0.02) & (5618; 4.43; -0.01)& (6239; 4.44; 0.07) & (6666; 4.49; 0.04) \\  
                        \cmidrule{2-5}        
                                $ \Delta T_{\rm eff} = \pm$ 50K  & $\pm0.03$ & $\pm0.03$& $\pm0.06$ & $\pm0.04$\\  
                        \hline
                        &     & & &\\
                        & HD 159868 & HD 168746 &HD 290327 & HD 41087 \\
                        & (5552; 3.91; -0.09) & (5561; 4.31; -0.11) & (5505; 4.41; -0.14) &(5562; 4.52; -0.13) \\ 
                        \cmidrule{2-5}                    
                        $ \Delta \rm log g$ = $\pm$ 0.1 dex    & $\mp0.02$ & $\mp0.01$& $\pm0.00$ &$\pm0.00$\\
                        \hline
                        &     & & & \\
                        
                        & HD 68089  & HD 161098 & HD 206163 & HD 203384\\
                        & (5597; 4.53; -0.77) & (5574; 4.49; -0.26)& (5506; 4.42; 0.02) & (5564; 4.42; 0.26) \\                    
                        \cmidrule{2-5}   
                        $ \Delta \rm([Fe/H]) =\pm$ 0.1 dex  & $\mp0.13$     & $\mp0.02$&$\mp0.02$&$\mp0.02$ \\
                                                           
                        \hline                
                \end{tabular}}
        \end{table*}


Since the discovery of the first exoplanet {around a main-sequence star} back in 1995 \citep{mayor}, the number of known exoplanets
has increased exponentially. Such discoveries have been made for a wide diversity of stars.
Many studies of the hosts stars have proved that the formation of high-mass planets
 correlates with the metal content of the star \citep[see][]{santos01, santos04,fischer05, sousa11a,
 mortier13, buchhave}. Moreover, it has been shown that the formation of planets at low metallicities
 is favoured by enhanced $\alpha$-elements \citep{adi_over, adi12c}.

\begin{figure}
        \centering
                \scalebox{1.3}[1.2]{
                        \includegraphics[angle=180,width=0.7\linewidth]{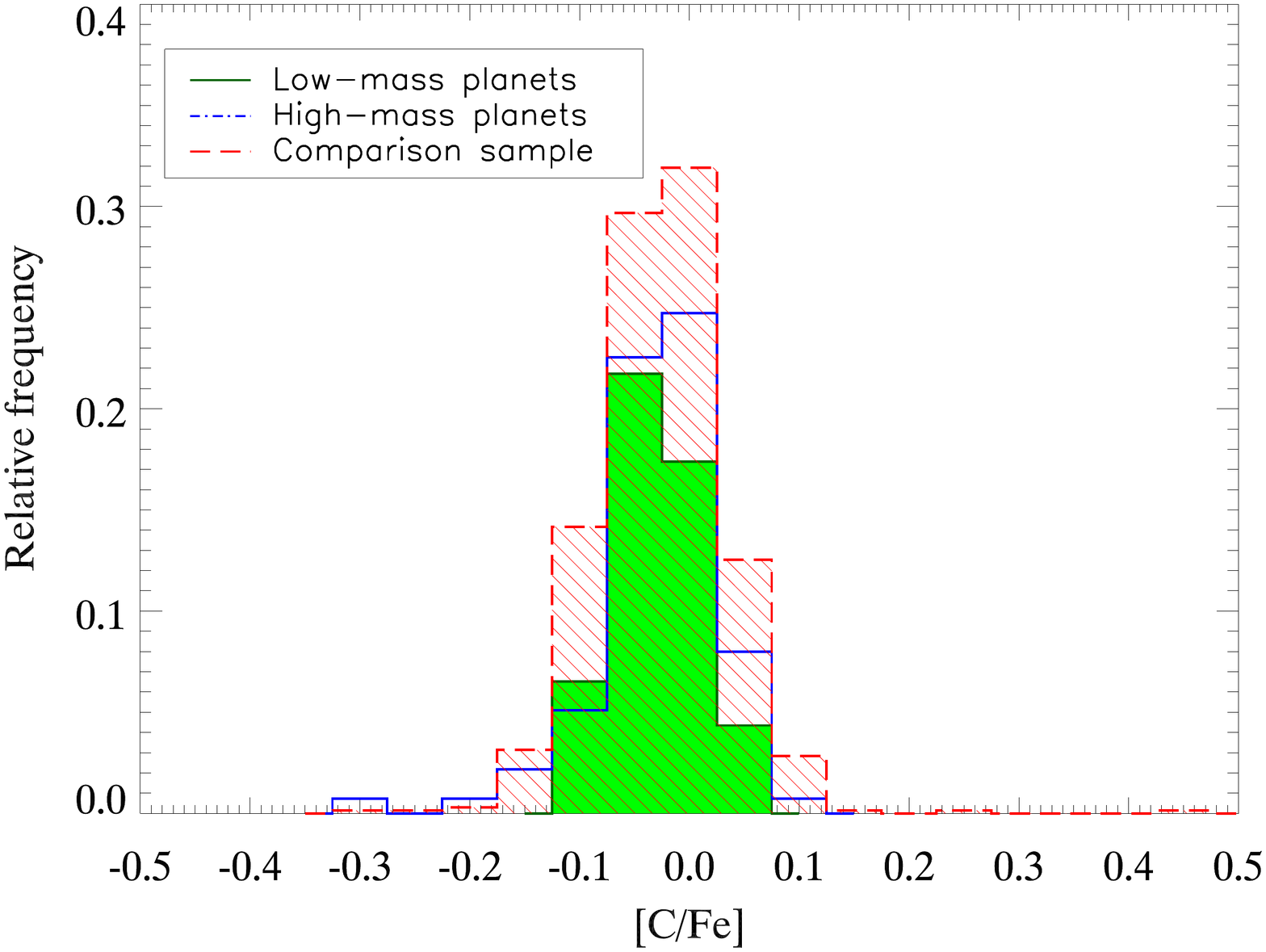}}

        \scalebox{1.3}[1.2]{
                \includegraphics[angle=180,width=0.7\linewidth]{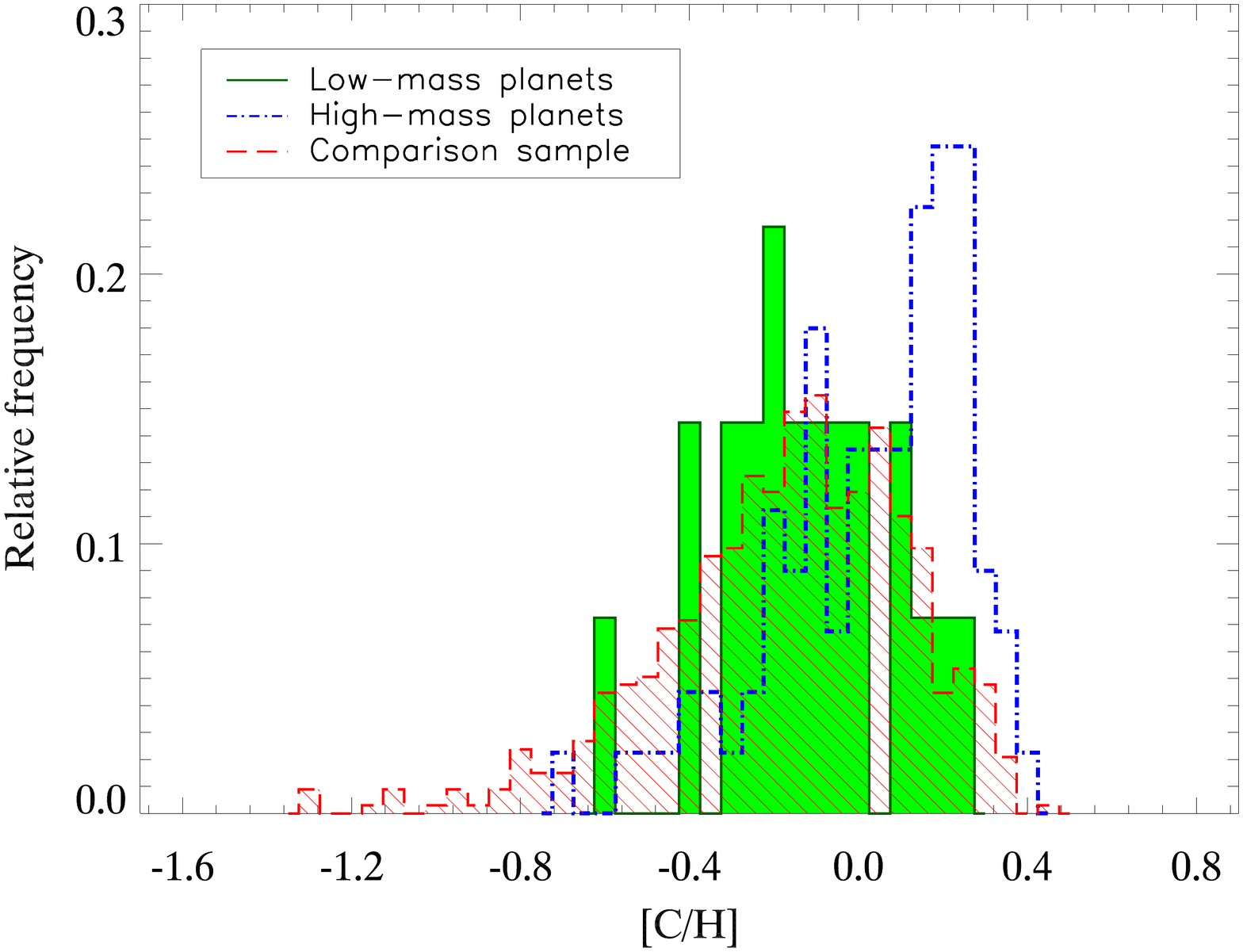}}
        \scalebox{1.3}[1.2]{
                \includegraphics[angle=180,width=0.7\linewidth]{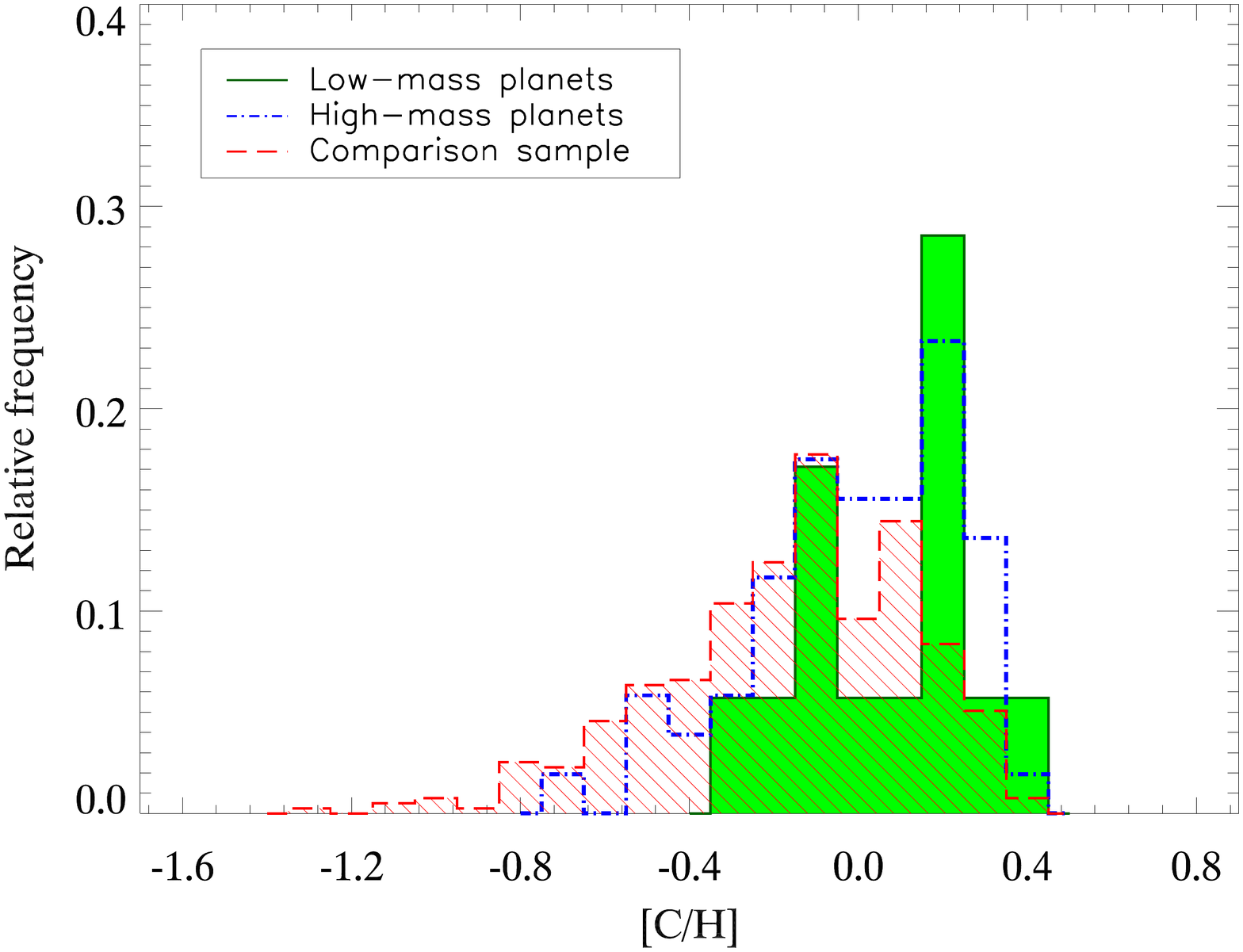}}

        \caption{{Top panel: [C/Fe] distributions of three different samples: stars without planets 
                (NP, in red), low-mass planets (LMP, in green), and high-mass planets (HMP, in blue). Middle panel:
                 [C/H] distributions for the three mentioned samples. Bottom panel:  [C/H] distributions for the same three groups, but considering only solar analogues ($T_{\rm eff} = T_{\rm eff, \odot} \pm 300$ K)}}
        \label{Fig:planet_histo}
\end{figure}

 In Fig. \ref{Fig:planet_histo} we can see the relative frequency of the {planetary samples} compared with the
 {comparison sample} (NP). The reader should be aware that undetected low-mass planets might be present 
 in the comparison sample. In the top panel we show the [C/Fe] distributions for all the samples and 
 find no differences among them. In the middle panel, we show [C/H] distributions for the same samples. 
 There is a slight offset between the samples, especially between the comparison 
and the high-mass planet sample. We can confirm that the high-mass planet offset found for [C/H] is due to the metal-rich 
nature of their host stars, as no carbon enhancement is found (see
Fig. \ref{Fig:met1}, bottom){. We found a completely different behaviour for the low-mass planetary sample,}
which does not seem to be preferentially metal-rich (\cite{ghezzi, sousa08,mayor11, sousa11b, buchhave, buchhave-latham}).
In the bottom planel, we show [C/H] but for solar analogue (stars with
($T_{\rm eff} = T_{\rm eff, \odot} \pm 300$ K, but not neccesarily solar gravity
or metallicity). Statistics for these results are provided in Table \ref{stats}.

\begin{table}[!t]
        \caption{{Statistics for the three samples studied (comparison sample - NP, low-mass 
        planets - {LMP}, high-mass planets - HMP) for [C/H] and [C/Fe] abundances. Also, [C/H] 
        abundances for those stars considered as solar analogues ($T_{\rm eff} = T_{\rm eff, \odot} \pm 300$ K) are provided. }}             
        \label{stats}      
        \centering   
        \scalebox{0.9}[1.]{      
                \begin{tabular}{ c c c c c c } 
                        \hline
                        \hline
                        & & &\multicolumn{1}{c}{[C/Fe]}\\
                        \cmidrule{2-6} 
                        &     & & & \\
                        & Mean &  SD & Median & Minimum & Maximum \\
                        \cmidrule{2-6}        
                        NP  & -0.03 & 0.06 & -0.02 & -0.31 & 0.44\\
                        LMP  & -0.03 & 0.04 & -0.03 & -0.11 & 0.05\\
                        HMP  & -0.03 & 0.06 & -0.02 & -0.29 & 0.12\\    
                        \cmidrule{1-6}                    
                        & & &\multicolumn{1}{c}{[C/H]}\\
                        \cmidrule{2-6} 
                        &     & & & \\
                        & Mean &  SD & Median & Minimum & Maximum \\
                        \cmidrule{2-6}        
                        NP  & -0.17 & 0.29 & -0.14 & -1.30 & 0.45\\
                        LMP  & -0.13 & 0.21 & -0.17 & -0.62 & 0.24\\
                        HMP  & 0.03 & 0.23 & 0.10 & -0.69 & 0.39\\    
                        \cmidrule{1-6}                    
                        & & &\multicolumn{1}{c}{$\rm [C/H]_{analogs}$}\\
                        \cmidrule{2-6} 
                        &     & & & \\
                        & Mean &  SD & Median & Minimum & Maximum \\
                        \cmidrule{2-6}        
                        NP  & -0.03 & 0.06 & -0.02 & -0.31 & 0.44\\
                        LMP  & -0.03 & 0.04 & -0.03 & -0.11 & 0.05\\
                        HMP  & -0.03 & 0.06 & -0.02 & -0.29 & 0.12\\    
                        
                        \hline                
                \end{tabular}}
        \end{table}

To see if there is any resemblance between each sub-sample of planet host with the {non-planet host} sample, we performed a Kuiper test (also known as invariant Kolmogorov--Smirnov (K--S) test),
 instead of the usual K--S test. Although K--S is suitable in our case, its sensitivity is higher around
 the median value, neglecting the tails. The Kuiper test compares two cumulative distribution functions
 and obtains the sum of the maximum distances above and below $S_{N}(x)$, where $S_{N}(x)$ is the cumulative 
distribution function of the probability distribution from which a dataset with $N$ events is drawn. In our
 study, a zero value of the K--S test refers to datasets with no similarities among them. Unity refers to
 the maximum similarity that can be found for the compared datasets \citep{numerical, kirkman}.

\begin{figure}
        \centering
        \scalebox{1.1}[0.95]{
                \includegraphics[angle=0,width=0.8\linewidth]{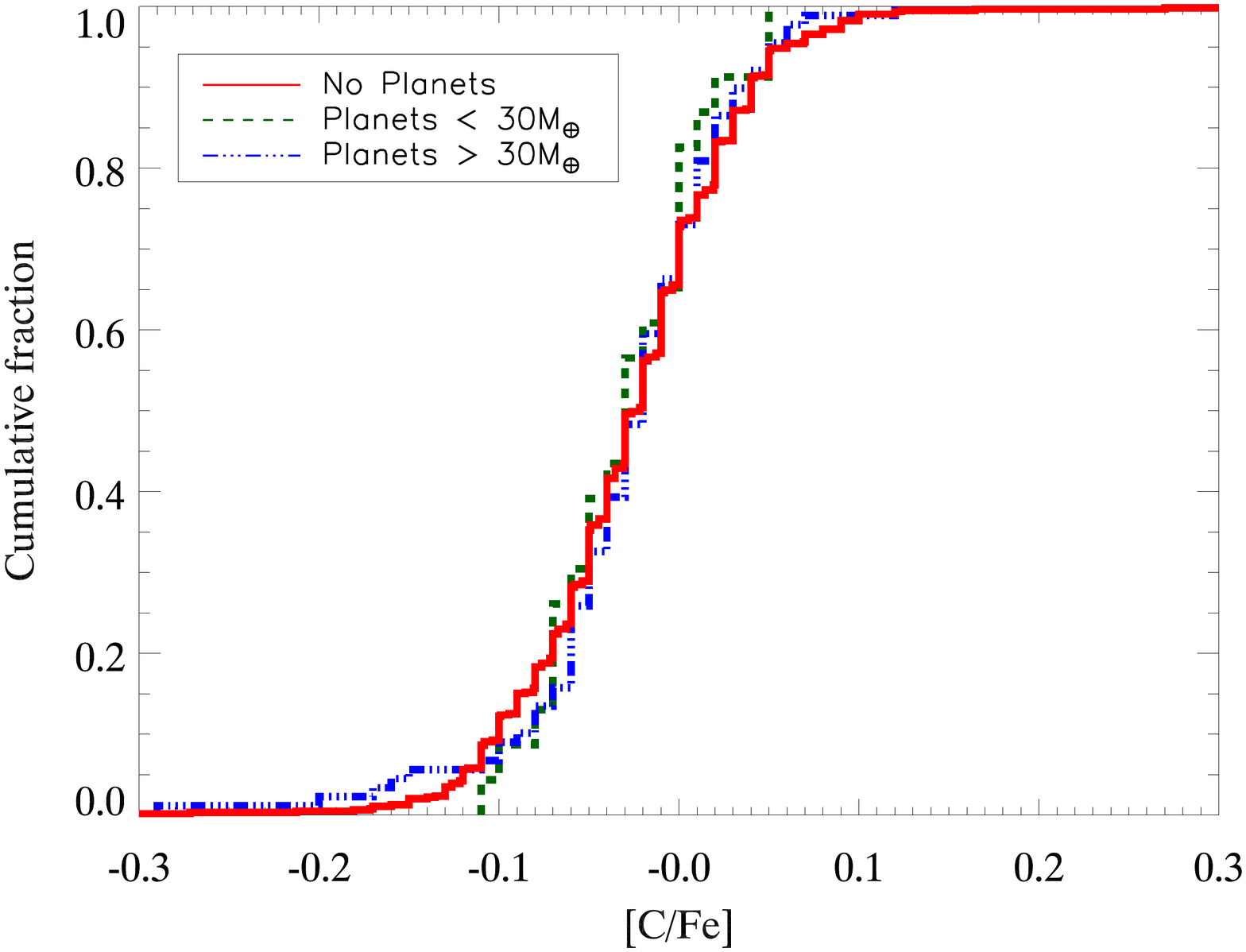}}
        \medskip
                \scalebox{1.1}[0.95]{
                \includegraphics[angle=0,width=0.8\linewidth]{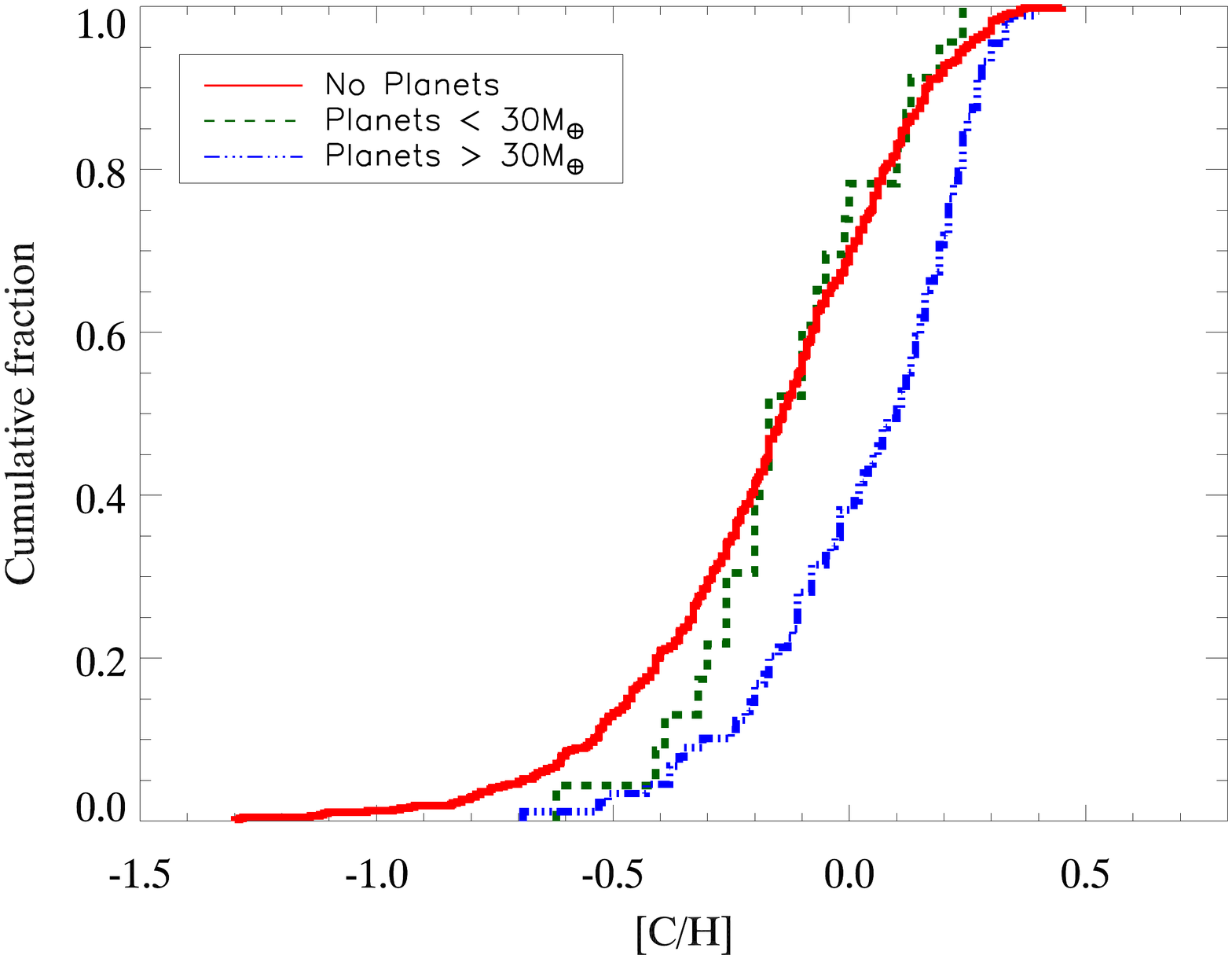}}
        \caption{Top: {Cumulative functions} for the three different samples (LMP, HMP, and NP) for [C/H] abundances. 
Bottom:  {Cumulative functions} for the three different samples (LMP - green, HMP - blue, and NP - red) for [C/Fe] abundances. }
        \label{Fig:kuiper}
\end{figure}

In Fig. \ref{Fig:kuiper} we can see the cumulative fraction for both [C/H] and [C/Fe] for the three cases: 
stars with high-mass planets (HMP) stars with low-mass planets (LMP), and the {comparison sample} (NP). In the top panel, we can see how all samples behave in a similar way. 
If we apply a Kuiper test to these samples, as shown in Table \ref{table:kuiper}, we can see that all the samples share 
similarities: when taking [C/Fe] into account, the high-mass planetary sample {behaves like the 
non-planet hosts and low-mass samples, as there is a probability of similarity between them.} The planetary 
samples, however,  separate  from each other in [C/H]. Whereas the 
small-planet sample is more in agreement with the {non-planet host} sample, the high-mass planet sample detaches itself
completely from this behaviour. {These results are consistent with the results represented in Fig. \ref{Fig:planet_histo}.}

\begin{figure}
        \centering
        \scalebox{1.3}[1.25]{
                \includegraphics[angle=0,width=0.7\linewidth]{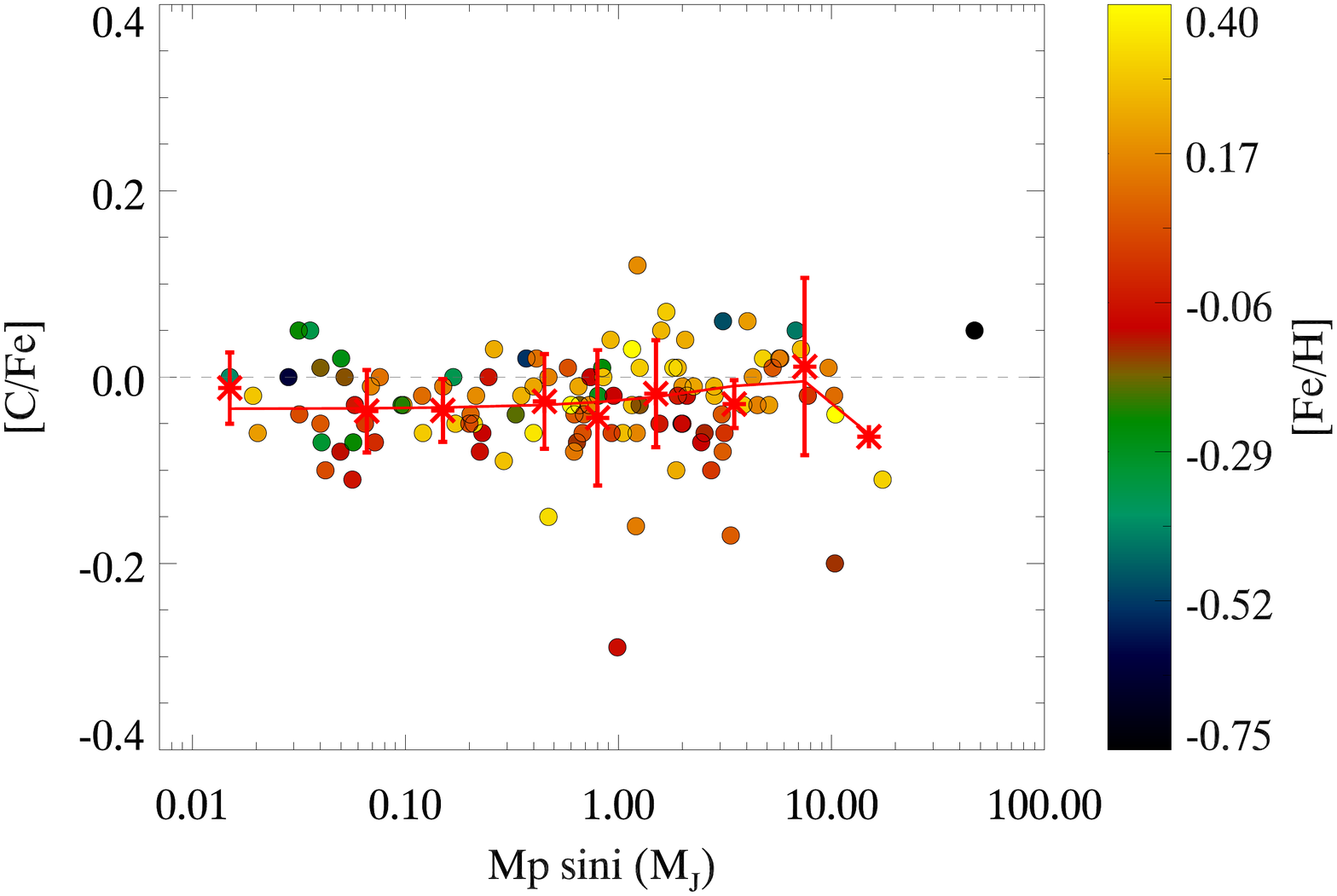}}
        \scalebox{1.3}[1.25]{
                \includegraphics[angle=0,width=0.7\linewidth]{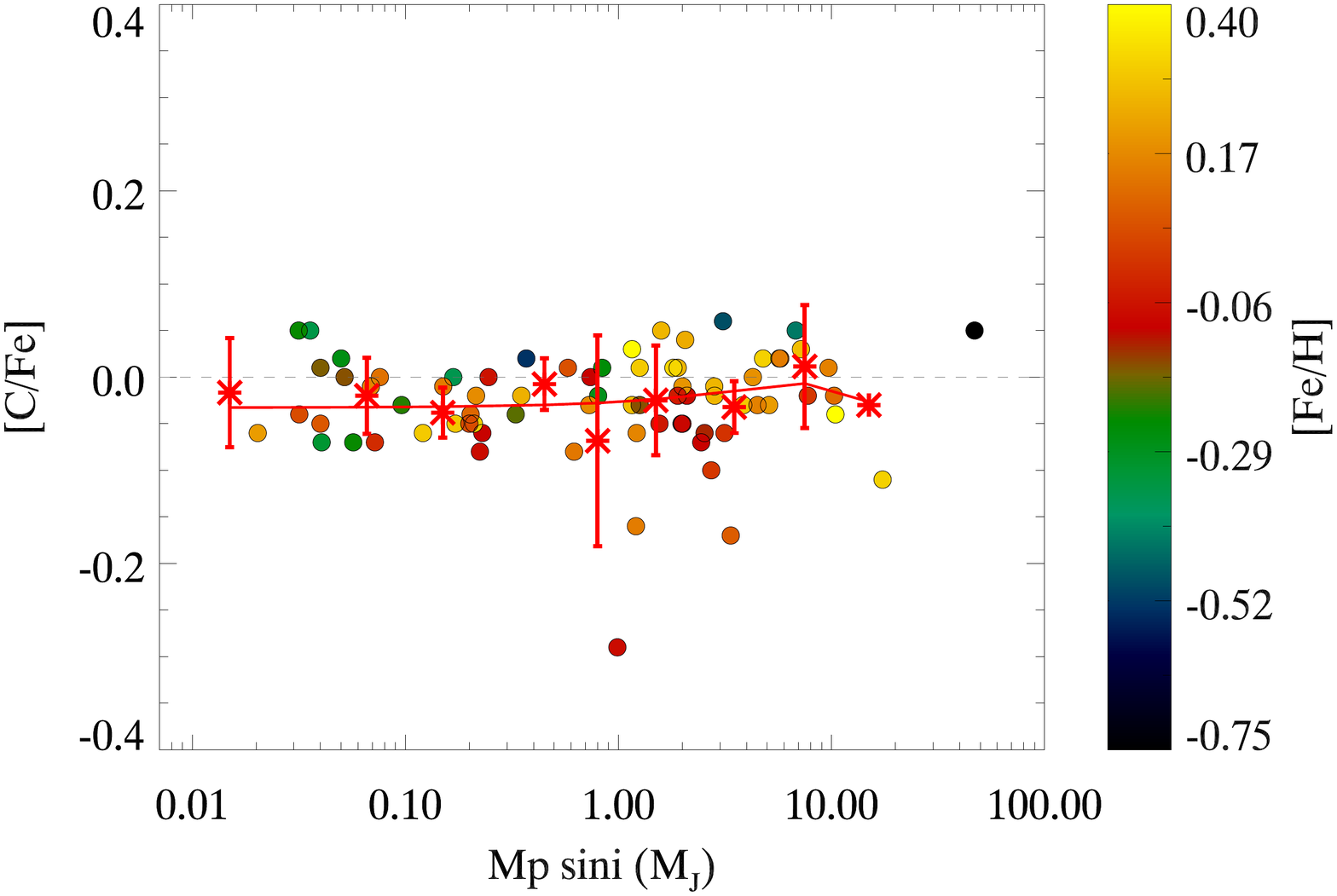}}
        \caption{[C/Fe] versus $M_{\rm p} sini$ plot. {Filled circles} represent the whole
                sample, whereas {red asterisks} represent binned values. Hotter colours represent 
                metallicities above solar while cooler colours represent metallicies below solar. Also in 
                red, the fit for these binned values. In the top panel, for all the studied sample, 
                in the bottom, for solar analogues ($T_{\rm eff} = T_{\rm eff, \odot} \pm 300$ K)}
        \label{planet_mass}
\end{figure}

We looked deeper into the relation between [C/Fe] and the mass of the planet, as shown in the top panel of Fig. \ref{planet_mass}.
The masses of the planets range between 0.0097 and 47 $M_{\rm J}$ (Only one star has a companion of 47 $M_{\rm J}$, which can be considered a brown dwarf, according to the standard definition of brown dwarf), but most stars have planets with less 
than 1 $M_{\rm J}$ orbiting around them. To remove this bias due to the lack of planets with higher masses, we
created bins with increasing steps, with bin sizes of 0.03, 0.07, 0.2, 0.3, 0.4, 2.0, 3.0, 5.0, 10.0, and 
27.0 $M_{\rm J}$. Error bars indicate the standard deviation of each bin. We can see how the [C/Fe] ratio shows 
a flat tendency for all masses because the deviation is negligible. We tested a possible relation between [C/Fe] 
and the planetary mass and obtained a covariance $s_{x,y}=0.001$, where $x=M_{\rm J}$ and $y=$ [C/Fe].
In the bottom panel of Fig. \ref{planet_mass}, we reduce the sample to solar analogues. {In our case, we considered
        as solar analogues those stars with $T_{\rm eff} = T_{\rm eff, \odot} \pm 300$ K.} We obtained a covariance $s_{x,y}=0.003$.

\section{Kinematics properties and stellar populations}

To study the kinematic properties of the sample stars and the stellar populations
they belong to{, we took the classification obtained by Adibekyan et al. (2012b), who applied both purely chemical 
        \citep[e.g.][]{adi11, adi13, recio14} and kinematic approaches
        \citep[e.g.][]{bensby, reddy} to classify their stars}. The Galactic space velocity components of the stars
were calculated in Adibekyan et al. (2012b) using the astrometric \footnote{The 
SIMBAD Astronomical Database (http://simbad.u-strasbg.fr/simbad/) was used \citep{wenger}.} 
and radial velocity data of the stars. The average errors in the $U$, $V$, and $W$ 
velocities are about 2--3 km s$^{-1}$. {The main source of the parallaxes and 
proper motions {was} the updated version of the Hipparcos catalogue \citep{leeuwen}.
Data for eight stars with unavailable Hipparcos information were
taken from the TYCHO Reference Catalog \citep{hog}.}

\begin{figure}
        \begin{center}
                
                \scalebox{1.3}[1.5]{
                        \includegraphics[angle=180,width=0.8\linewidth]{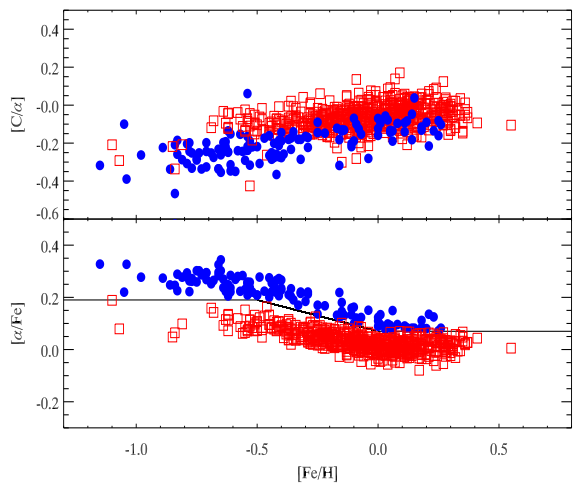}}
                
                \caption{[C/$\alpha$] and [$\alpha$/Fe] vs.\ [Fe/H]. Stars that are enhanced 
                        in $\alpha$-elements are shown as blue squares and non-$\alpha$-enhanced stars are represented by red {squares}.}
                \label{alfa}
        \end{center}
        
\end{figure}

The separation of the Galactic stellar components based only on stellar abundances 
is probably superior to that based on kinematics alone \citep[e.g.][]{navarro11,
        adi11} because chemistry is a relatively more stable property of 
sunlike stars than their spatial positions and kinematics. 
We used the 
position of the stars in the [$\alpha$/Fe]--[Fe/H] plane {(here $\alpha$ refers to the average 
abundance of Si, Mg, Ti)} to separate the thin- and thick-disc stellar components. {We recall that Ca was not included in the $\alpha$ index, because at super-solar metallicities the [Ca/Fe] trend differs from that of other $\alpha$-elements \citep{adi12}}. 
We have adopted the
boundary (separation line) between the stellar populations from \citet{adi11}. According to our separation, 640 stars 
($\approx$85 per cent) are not enhanced in $\alpha$-elements and belong to the thin-disc population. The [$\alpha$/Fe] versus [Fe/H] plot 
for the sample stars is shown in the bottom panel of Figure \ref{alfa}.

In the top panel of Fig. \ref{alfa} we show the dependence of [C/$\alpha$] on the metallicity. This trend is expected if the
C abundance scales with the iron abundance, as was suggested above. It is interesting to note that 
at super-solar metallicities, [C/$\alpha$] remains almost constant for both samples.

        \begin{table}
                
                \caption{{Kuiper test} for the sample. {We} tested the relation between the low-mass planets 
                        (LMP) sample, the high-mass planets sample (HMP) and the {comparison sample} (NP) sample, for both [C/Fe] and [C/H].
                        Zero represents no similarity between the samples while unity represents the maximum similarity.}
                
                \begin{center}
                        \scalebox{0.9}[1]{
                                \begin{tabular}{l c c}
                                        \hline
                                        \hline
                                        
                                        Sample& [C/Fe]& [C/H]\\
                                        \midrule
                                        LMP - NP & 0.54 & 0.72 \\
                                        HMP - NP & 0.65  & 2.19e-08 \\
                                        LMP - HMP & 0.85 & 3.78-03 \\                   
                                                \bottomrule
                                \end{tabular}}
                                \label{table:kuiper}
                        \end{center}
                \end{table}

\begin{figure}[!h]
        \centering
        \scalebox{1.3}[1.4]{
                \includegraphics[angle=180,width=0.8\linewidth]{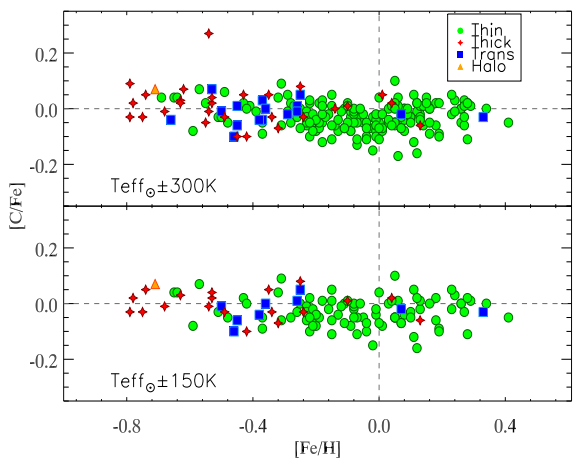}}
        \caption{[C/Fe] versus [Fe/H] for solar analogues with S/N $\ge$ 200. Filled green dots refer to 
                to the thin-disc population, filled blue squares to transition objects, orange triangles to halo stars, 
                and red stars to thick-disc stars.}
        \label{pop}
\end{figure}

In Fig. \ref{pop} we present [C/Fe] against [Fe/H] {for the sample of solar analogues}. The selected stars
have $T_{\rm eff} = T_{\rm eff, \odot} \pm 300$ K in the top panel and
$T_{\rm eff} = T_{\rm eff, \odot} \pm 150$ K in the bottom panel. These solar 
analogues also have S/N > 200 to ensure optimal accuracy in our conclusions.

Figure \ref{pop} shows that, opposite to the chemical separation (using Ti, Mg, and Si for 
the $\alpha$ abundances), the thin and thick discs are not chemically different for C. In \cite{adi12}, 
it was suggested that several $\alpha$-elements show different trends with metallicity.

\section{Summary and conclusions}

We present carbon abundances for 1110 solar-type stars observed with HARPS. In our sample, 143  of
 the 1110 are planet-hosts and 967 stars  have no known planetary companion. All the targets have been
 studied using spectral synthesis of the CH band at 4300\AA. All the stars within our sample have effective
 temperatures between 4400 K and 7212 K, metallicities from $-1.39$ to $0.55$ dex, and surface gravities from $3.59$ to $4.96$ dex.

We have performed a detailed spectral analysis of this sample to obtain precise carbon abundances and investigate
 possible differences between stars with planets and the {comparison sample}. We confirm that both samples,
 planet host stars and comparison stars, show no different behaviour when we study their carbon 
abundances as a function of different stellar parameters such as $T_{\rm eff}$ and [Fe/H].

 We have compared our results with similar work \citep{elisa10}, as seen in Fig \ref{comp}. When
 comparing stars common to both works, we get agreement between the atomic and the molecular
 results, thus ensuring that molecular abundances from the CH 4300\AA \space band can be a reliable alternative to atomic values.

We searched for a correlation between the presence of planets and carbon abundances. We found that {both planet-hosts and non-planet-hosts
 usually have [C/Fe] $\le$ 0.0 {($\sim$65\% of the sample). We searched for a correlation between the 
carbon abundance and the planetary masses but found a flat trend for all the masses.} 

We looked for similarities between the three samples studied: the {non-planet host} samples and the low- and high-mass 
planetary sample. We found that both low- and high-mass planet hosts are not different 
from the {comparison sample}. None of the samples show signs of carbon enhancement.

{We performed a chemical and kinematic separation for our sample and obtained a clear
separation between the thin and the thick disc populations based on a dependence of [C/$\alpha$] with metallicity.
        However, no differences in carbon abundances can be found for the different stellar 
        populations.}
{We conclude that the molecular CH band located at 4300\AA \space can be used to obtain 
reliable and accurate carbon abundances and therefore provide an accurate measurement of C/O 
ratios that can be used to investigate the mineralogy of exoplanets (Su\'arez-Andr\'es et al. 2016, in prep).}

\begin{acknowledgements}
        JIGH acknowledges financial support from the Spanish Ministry of Economy and Competitiveness (MINECO) under 
the 2013 Ram\'on y Cajal programme MINECO RYC-2013-14875, and the Spanish ministry project MINECO AYA2014-56359-P. V.Zh.A., E.D.M., N.C.S. and S.G.S. acknowledge the support from Funda\c{c}\~ao para a Ci\^encia e a Tecnologia (FCT) through national funds 
and from FEDER through COMPETE2020 by the following grants UID/FIS/04434/2013 \& POCI-01-0145-FEDER-007672, PTDC/FIS-AST/7073/2014 \& POCI-01-0145-FEDER-016880 and PTDC/FIS-AST/1526/2014 \& POCI-01-0145-FEDER-016886. V.Zh.A., N.C.S. and S.G.S. also acknowledge the support from FCT through Investigador FCT contracts IF/00650/2015, IF/00169/2012/CP0150/CT0002 and IF/00028/2014/CP1215/CT0002; and E.D.M. acknowledges the support by the fellowship SFRH/BPD/76606/2011 funded by FCT (Portugal) and POPH/FSE (EC). This work has made use of the VALD database, 
operated at Uppsala University, the Institute of Astronomy RAS in Moscow, and the University of Vienna.

\end{acknowledgements}


        
\Online
\end{document}